\documentclass{PoS}

\title{Prospects for Precision Higgs Physics at Linear Colliders}

\ShortTitle{Prospects for Precision Higgs Physics at Linear Colliders}

\author{\speaker{Frank Simon}\\
       {\rm On behalf of the Linear Collider Physics \& Detector Studies}\\
       Max-Planck-Institut f\"ur Physik, Munich, Germany\\
       E-mail: \email{fsimon@mpp.mpg.de}}

\abstract{A linear $e^+e^-$ collider provides excellent possibilities for precision measurements of the properties of the Higgs boson.  At energies close to the Z-Higgs threshold, the Higgs boson can be studied in recoil against a Z boson, to obtain not only a precision mass measurement but also direct measurements of the branching ratios for most decay modes, including possible decay to invisible species.  At higher energies, the Higgs boson coupling to top quarks and the Higgs boson self-coupling can also be measured.  At energies approaching 1 TeV and above, the rising cross section for Higgs production in WW fusion allows the measurement of very small branching ratios, including the branching ratio to muon pairs.  These experiments make it possible to determine the complete profile of the Higgs boson in a model-independent way. The prospects for these measurements are summarized, based on the results of detailed simulation studies performed within the frameworks of the CLIC conceptual design report and the ILC technical design report.}

\FullConference{36th International Conference on High Energy Physics\\
		 4-11 July 2012\\
		 Melbourne, Australia}

\begin{document}

\section{Introduction}

The recent discovery of a new heavy particle, which is so far consistent with the Standard Model Higgs boson \cite{:2012gk, :2012gu} has substantial consequences for our understanding of the structure of matter, calling for a detailed investigation of the properties of this particle. It is crucial to precisely establish its mass and its quantum numbers, its coupling to fermions, bosons and its self-coupling, and to determine if it is the single, fundamental scalar of the Standard Model or if it is part of a more extended Higgs sector, or a composite state bound by so-far unknown interactions. The mass of around 125 GeV provides for a large number of final states with accessible branching fractions, allowing detailed investigations of the mass dependence of the couplings. While the LHC is expected to provide decisive answers on some of the questions outlined above in the coming years, a full exploration of this newly discovered sector of particle physics will not be possible with the LHC alone. A linear electron-positron collider operated at several different energies provides substantial additional precision in most areas of Higgs physics and is capable of fully model-independent measurements of the couplings. This increased precision will allow accurately identifying possible non-Standard Model Higgs sectors, which may manifest themselves in deviations of the couplings, which are expected at the percent level for gauge bosons and at the few 10\% level for fermions in typical two-Higgs-doublet models \cite{Gupta:2012mi}.

Two concepts for a linear collider are currently being developed, based on different acceleration schemes which result in different energy reaches. The International Linear Collider (ILC) \cite{BrauJames:2007aa} is based on superconducting RF structures while the Compact Linear Collider (CLIC) \cite{Lebrun:2012hj} uses normal-conducting two-beam acceleration technology. For the ILC, the technical design report is currently being finalized, while for CLIC the conceptual design report has recently been completed. The ILC is planned as a 500 GeV machine, possibly with first lower-energy stages, and an upgrade path up to 1 TeV. The higher acceleration gradient of the CLIC technology plans to reach up to 3 TeV, with the machine foreseen to be implemented in stages to maximize the physics potential. The detector concepts being developed for a linear collider are based on particle flow event reconstruction with precise tracking and vertexing to fulfill the requirements imposed by the physics goals of the project. Two detector concepts, ILD \cite{Abe:2010aa} and SiD \cite{Aihara:2009ad} are being developed for the ILC, and corresponding variants for CLIC, accounting for the higher energy and the more challenging background conditions, have been established \cite{Linssen:2012hp}.

In the following, the Higgs physics program at a Linear Collider at various energies will be outlined, naturally divided by the dominating production modes in a region below a center-of-mass energy of 500 GeV and a region above. The discussion is independent of the detailed choice of the machine technology or detector concept. Most results presented here are based on detailed detector simulations performed in the context of the ILC and CLIC physics and detector studies, and are typically assuming a Higgs mass of 120 GeV, and are taken from recent reports \cite{Lebrun:2012hj, Abe:2010aa, Aihara:2009ad, Brau:2012hv, Ono:2012ah} as well as from ongoing studies in the framework of the ILC TDR. 

\section{Higgs Production at Linear Colliders}

\begin{figure}
\centering
\includegraphics[height=0.2\textwidth]{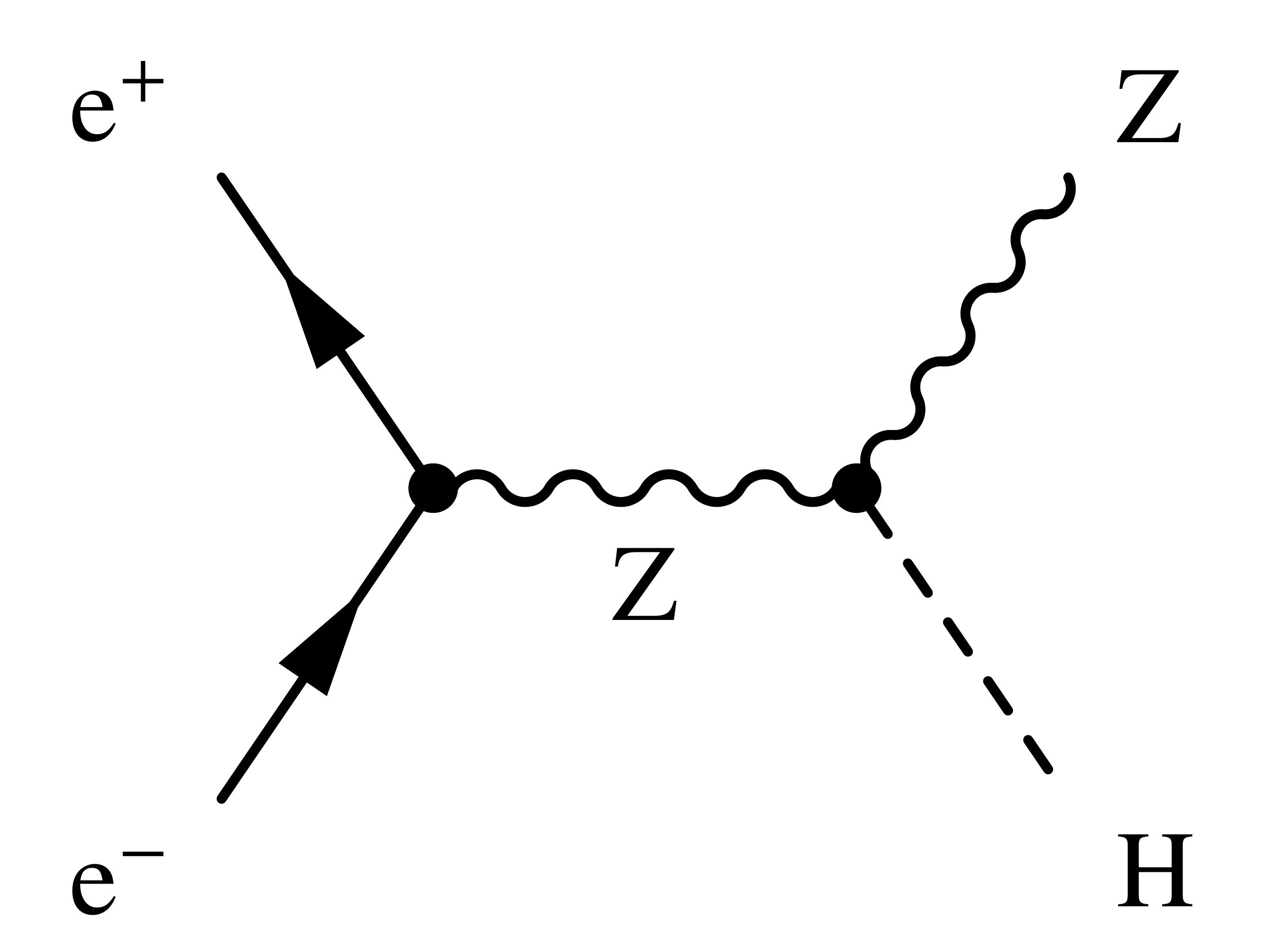}
\hspace{0.1\textwidth}
\includegraphics[height=0.22\textwidth]{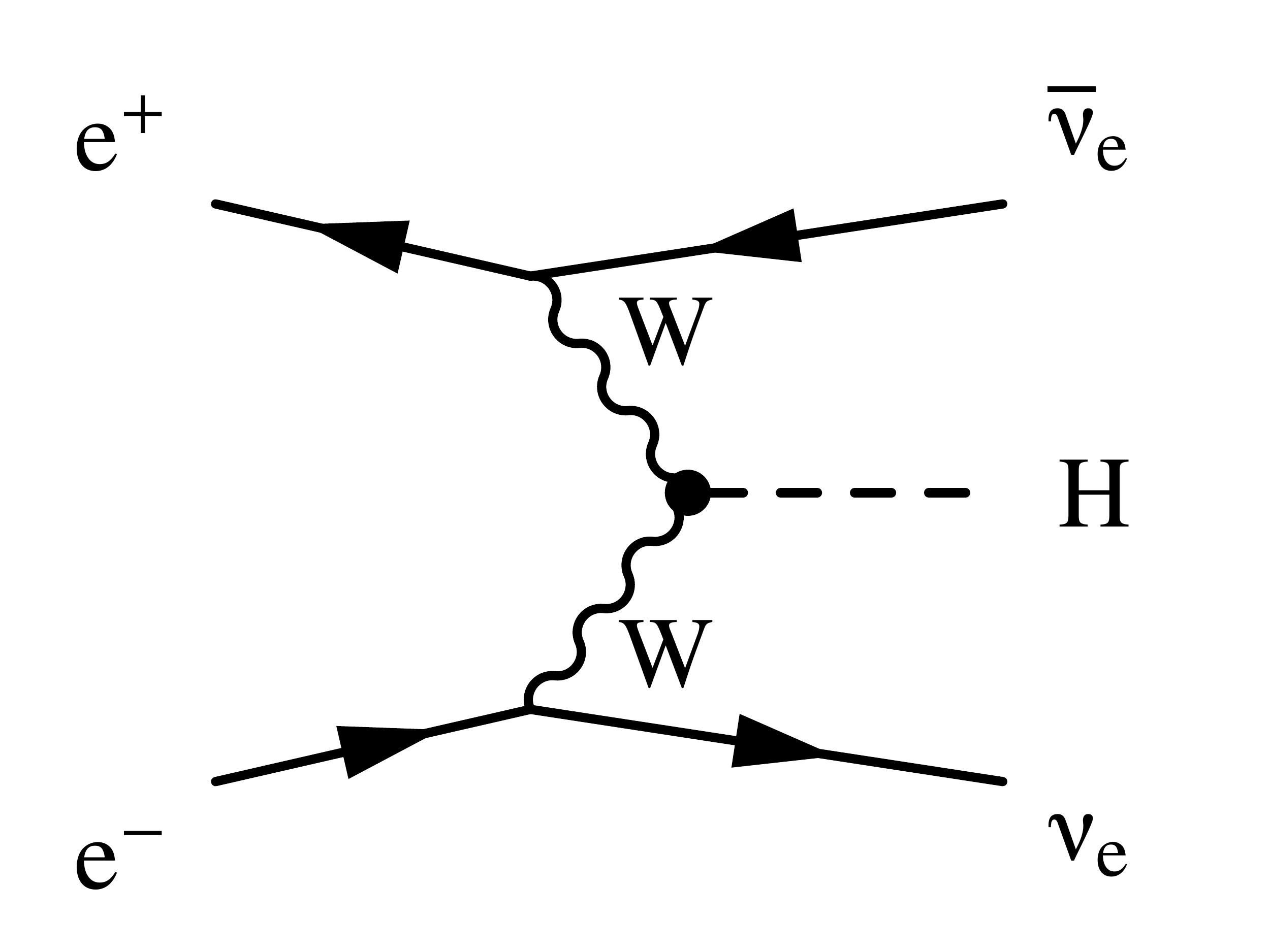}\\
\vspace{0.02\textwidth}
\includegraphics[height=0.2\textwidth]{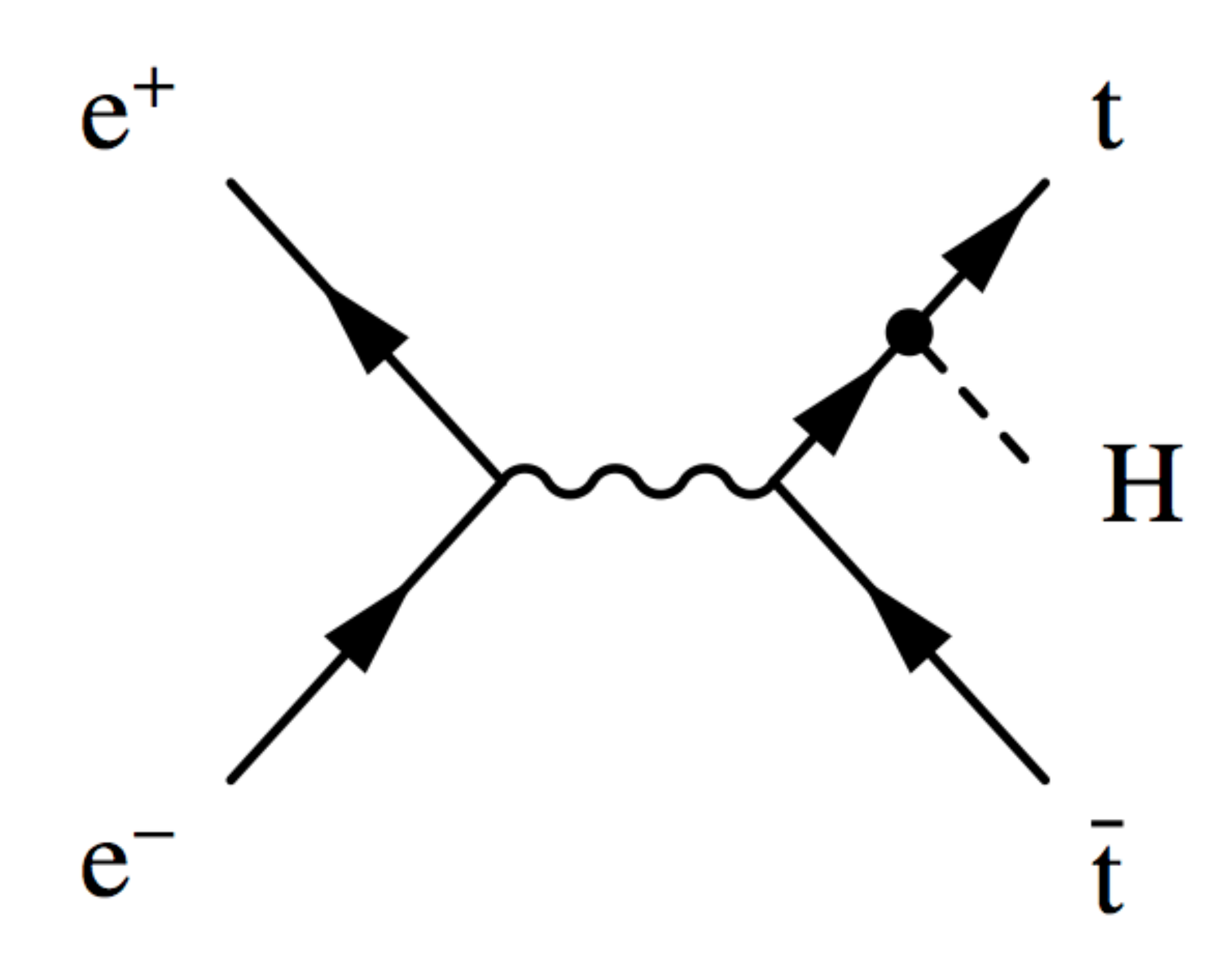}
\hspace{0.05\textwidth}
\includegraphics[height=0.2\textwidth]{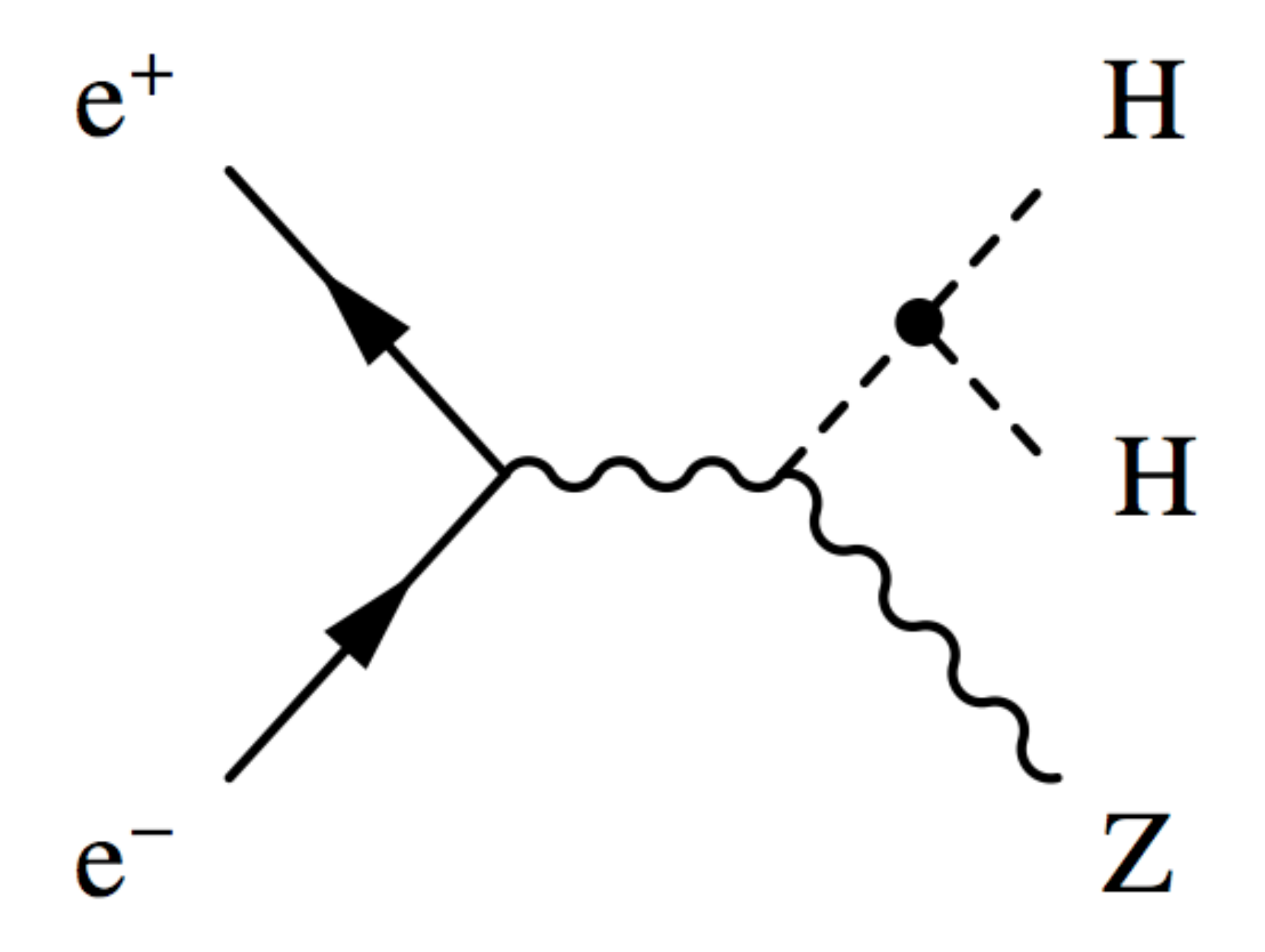}
\hspace{0.05\textwidth}
\includegraphics[height=0.2\textwidth]{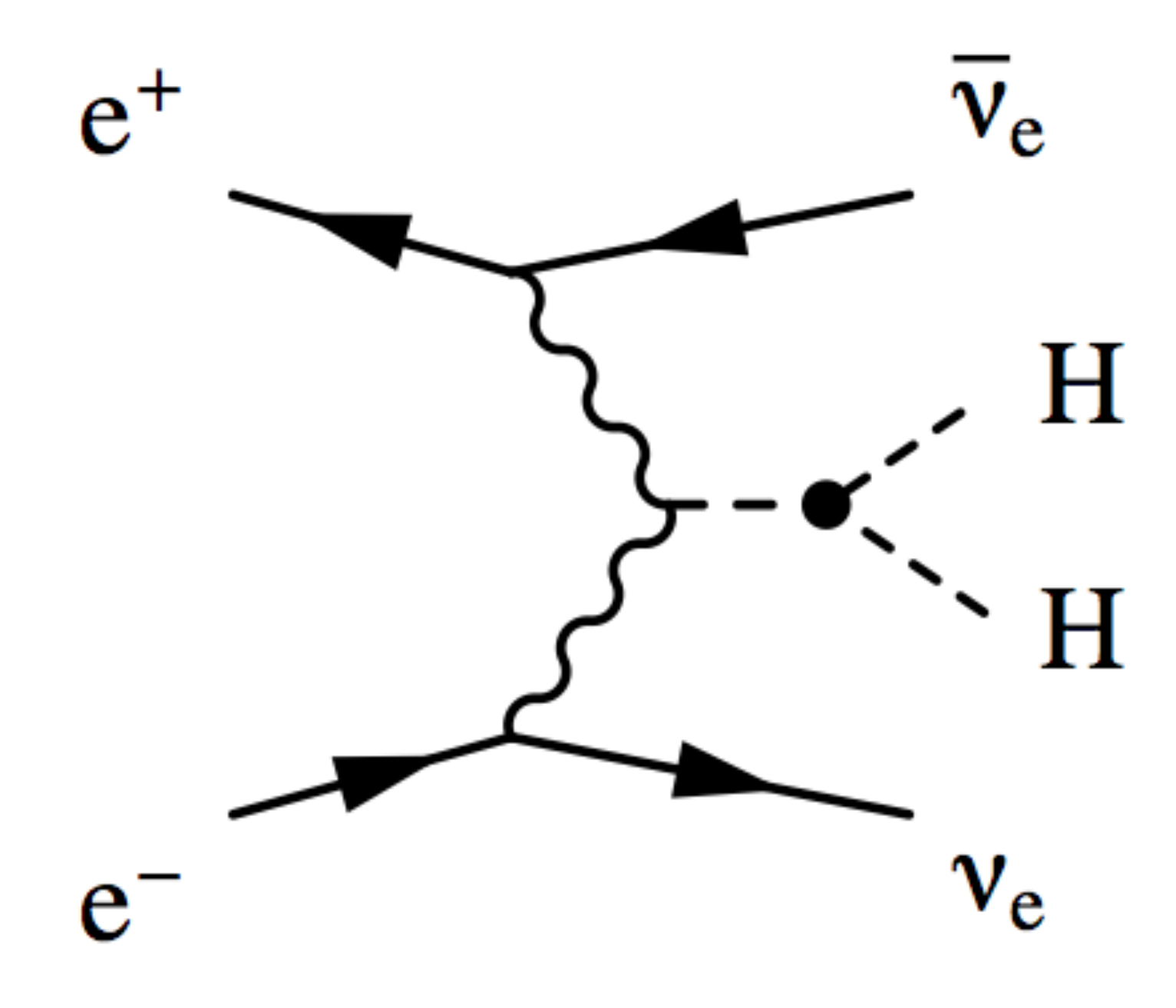}

\caption{Higgs production in $e^+e^-$ annihilation. Top Row: The two leading processes, Higgs-strahlung and vector boson fusion. Bottom row: Top-Higgs production and double Higgs production in the strahlungs- and the fusion process.}
\label{fig:HZ}
\end{figure}

\begin{figure}
\centering
\includegraphics[width=0.498\textwidth]{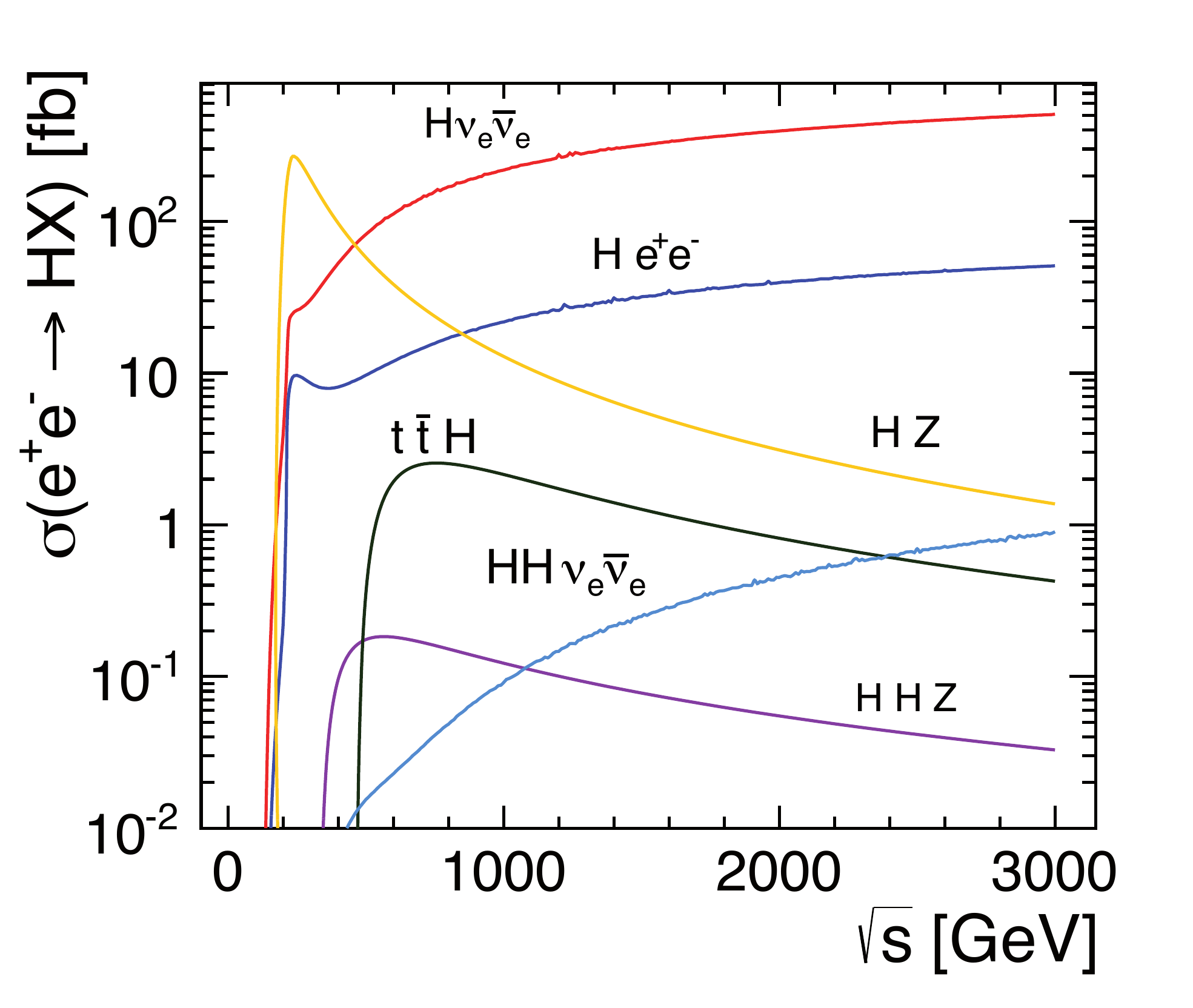}
\hfill
\includegraphics[width=0.441\textwidth]{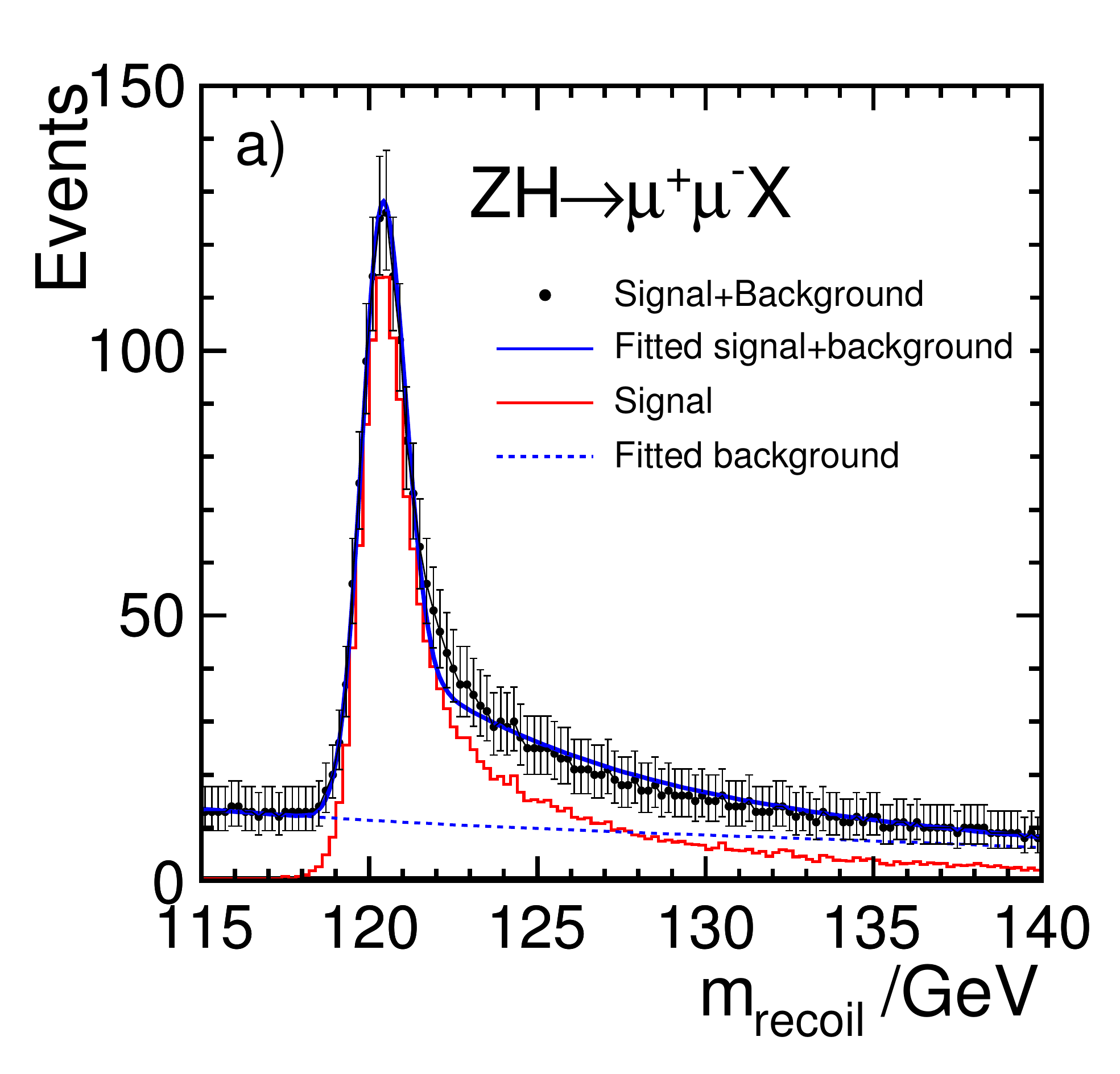}
\caption{Left: Tree-level cross sections for various Higgs processes as a function of $e^+e^-$ collision energy for unpolarized beams. Right: Recoil mass distribution for the process $e^+e^-\rightarrow \mathrm{ZH} \rightarrow \mu^+\mu^-X$ together with non-Higgs background at a center-of-mass energy of 250 GeV for $m_\mathrm{H}$ = 120 GeV, scaled to an integrated luminosity of 250 fb$^{-1}$. The error bars show the expected statistical uncertainty at each point. Figure taken from \cite{Abe:2010aa}.}
\label{fig:CrossSections}
\end{figure}

Higgs bosons are produced in $e^+e^-$ collisions by two main processes, the s-channel Higgs-strahlung process where the Higgs boson is radiated off a Z boson, and the t-channel W boson fusion process, as shown in the top row of Figure \ref{fig:HZ}. LHC has provided evidence for the coupling of the new boson to both W and Z bosons \cite{:2012gk, :2012gu}, confirming that these production mechanisms are accessible. At low center-of-mass energies, the process $e^+e^-\rightarrow \mathrm{ZH}$ dominates, with a cross-section maximum at approximately 250 GeV. Since this cross-section falls rapidly with increasing energy, while the vector boson fusion cross-section increases logarithmically with energy, the \mbox{$e^+e^-\rightarrow \mathrm{H}\nu\nu$} process dominates at energies above approximately 450 GeV, as illustrated in \mbox{Figure \ref{fig:CrossSections} {\it left}}. Depending on energy and integrated luminosity, 10$^5$ to 10$^6$ Higgs bosons are expected to be produced at each energy stage, with the highest numbers reached at 3 TeV due to the high production cross-section combined with the high instantaneous luminosity of a multi-TeV linear collider.

At energies of 500 GeV and above, additional processes, such as $e^+e^-\rightarrow t\bar{t}\mathrm{H}$,  $e^+e^-\rightarrow\mathrm{ZHH}$ and $e^+e^-\rightarrow \mathrm{HH}\nu\nu$ become accessible, illustrated in the bottom row of  Figure \ref{fig:HZ}. As shown in Figure \ref{fig:CrossSections} {\it left}, these processes have substantially lower cross-sections than the leading production modes, and require high integrated luminosities for precise investigations.

In general, the use of polarized beams, which is possible and foreseen at linear colliders, provides  a substantial increase of the signal cross sections,  further enhancing the potential for Higgs physics. At 3 TeV, the signal and background cross section of double Higgs production are increased by a factor of two with electron (80\%) and positron (30\%)  polarization.

\section{Measurements below 500 GeV}


The relatively clean environment and the well-defined collision energy at a linear collider together with the precise momentum resolution provided by the detectors allow to make a model-independent measurement of the HZZ coupling in the Higgs-strahlung process. By reconstructing only the decay products of the Z boson with the highest precision given by Z$\rightarrow\mu^+\mu^-$, the cross-section of the ZH process can be determined through the recoil mass spectrum, as shown for the ILD detector at 250 GeV with an integrated luminosity of 250 fb$^{-1}$ in Figure \ref{fig:CrossSections} {\it right}. The distribution peaks at the Higgs mass, with the tail to higher masses due to the combined effects of initial state radiation and beamstrahlung on the center-of-mass energy of the $e^+e^-$ annihilation.  

At energies of 250 GeV and 350 GeV, a precision of 3\% to 4\% on the cross-section of the Higgs-strahlung process, translating to a precision of 1.5\% to 2\% on the coupling $g_\mathrm{HZZ}$, can be achieved. The slightly higher precision at 250 GeV is due to the higher resolution for lower-momentum muons and due to the higher cross-section, which is somewhat offset by the increased luminosity at 350 GeV.  With the same technique, possible invisible decays of the Higgs boson can be constrained to the 1\% level by considering hadronic decays of the Z boson in addition.

The explicit reconstruction of the Higgs boson in addition to the recoiling Z boson permits the measurement of the branching fractions of the decays into $b$ and $c$ quarks, $\tau$, WW$^*$ and gluons. 
For these measurements precise flavor tagging to separate $b$ and $c$ jets as well as light jets is crucial, in addition to excellent particle flow performance for the efficient identification of $\tau$ leptons and for the reconstruction of hadronic W decays. Due to more favorable background conditions and improved flavor tagging performance arising from higher boosts, a slightly higher precision is achieved at 350 GeV compared to 250 GeV for these measurements. Taking the model-independent measurement of $g_\mathrm{HZZ}$ as normalization, the different couplings can be determined directly from the measured branching ratios. The expected combined precision ranges from below 2\% for $g_{\mathrm{H}bb}$ to 3\% to 4\% for the other couplings. 
For the branching ratio of $\mathrm{H}\rightarrow gg$ a statistical precision of 7\% to 10\% is expected, which can not be directly transformed to a coupling, but provides model-dependent sensitivity to the coupling to the top quark through loop contributions.

In addition to the measurement of the couplings, the measurement of the Higgs mass is possible with a precision of 40 MeV or better, both by direct reconstruction of the full final state, and by the measurement of the recoil spectrum. Furthermore, the cross-section behavior close to the production threshold and angular correlations of the decay products in the Higgs-strahlungs-process are sensitive to the spin and quantum numbers and will yield a measurement of the spin and CP properties to a few percent.

\section{Measurements at 500 GeV and up into the multi-TeV region}

Energies of 500 GeV up to 1 TeV provide the possibility to directly measure the coupling of the Higgs to the top quark. QCD $t\bar{t}$ bound state effects enhance the cross section at 500 GeV over the simple tree-level expectation shown in Figure \ref{fig:CrossSections} {\it left}, making a measurement possible with polarized beams and integrated luminosities of 1 ab$^{-1}$. At 1 TeV with 1 ab$^{-1}$ and polarized electron and positron beams, currently ongoing studies indicate that a measurement of the $g_{\mathrm{H}tt}$ coupling below the 5\% level can be achieved. 

The measurement of the tri-linear self-coupling provides direct access to the Higgs potential, and is thus crucial to establish the Higgs mechanism experimentally. The two double-Higgs production processes, $e^+e^-\rightarrow\mathrm{ZHH}$ and $e^+e^-\rightarrow \mathrm{HH}\nu\nu$, are available for measuring the trilinear self-coupling, with the former reaching its cross-section maximum at a center-of-mass energy of 500 GeV, while the cross-section for the latter is dominating above \mbox{1.1 TeV} and increases towards higher energies. Due to the low production cross-sections, the high non-double-Higgs background levels and the complex final state, this is a challenging measurement also at a linear collider. Currently ongoing studies indicate that with integrated luminosities of 2 ab$^{-1}$ it will be possible to provide significant evidence for the self-coupling at 500 GeV, while the same integrated luminosity at 3 TeV will yield a measurement of the self-coupling $\lambda$ on the 15\% to 20\% level. Polarization of both beams (80\% for electrons, 30\% for positrons) increases the signal cross-section by a factor of two, which will improve the measurement further. 

Access to the WW fusion process at energies of 500 GeV and above provides the possibility to measure the total width of the Higgs boson, by giving an independent measurement of the HWW coupling in addition to the branching ratio measurement of $\mathrm{H}\rightarrow\mathrm{WW}^*$ discussed above. Exploiting $\Gamma_\mathrm{H} = \Gamma(\mathrm{H}\rightarrow\mathrm{WW}) / \mathrm{BR}(\mathrm{H}\rightarrow\mathrm{WW}^*)$, with $\Gamma(\mathrm{H}\rightarrow\mathrm{WW})$ determined from the measured WW fusion cross-section, the total width of the boson can be determined with a precision of 4\% to 5\%.

In addition to these measurements, which require energies of 500 GeV and above to be accessible, the high luminosity at a multi-TeV linear collider combined with the increasing cross-section of the WW fusion process yields the possibility for precise measurements of rarer Higgs decays. With an integrated luminosity of 2 ab$^{-1}$ at 3 TeV, the coupling to charm quarks, $g_{\mathrm{H}cc}$, can be measured with a statistical precision of 2\%, and a measurement of the coupling to muons, $g_{\mathrm{H}\mu\mu}$, becomes possible at the 7.5\% level. At this energy, the measurement of the ratio of the WW and the ZZ fusion processes also enables a measurement of the ratio of $g_\mathrm{HWW}/g_\mathrm{HZZ}$ with sub-percent precision.

\section{Summary}

\begin{figure}
\centering
\includegraphics[width=0.45\textwidth]{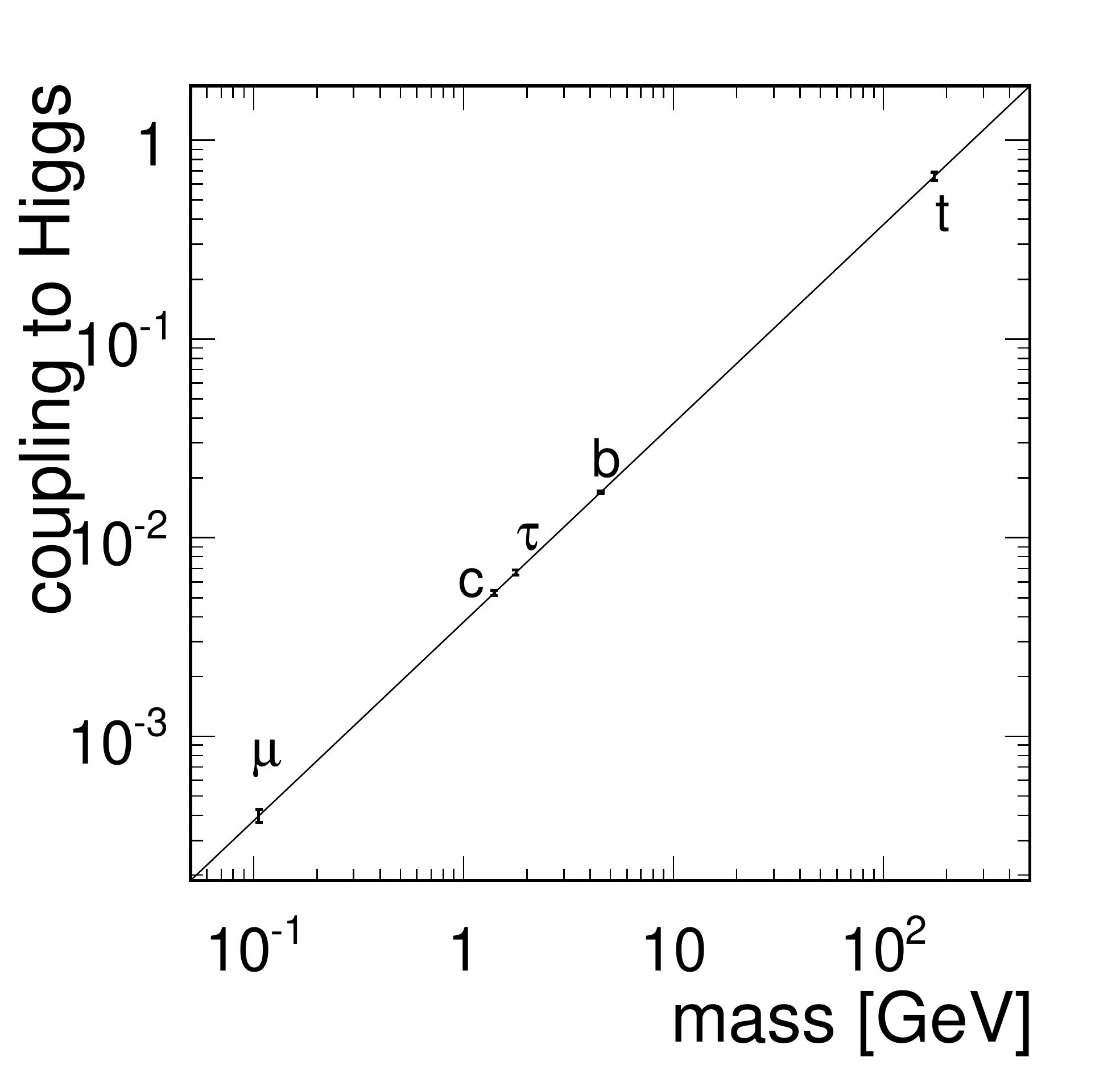}
\hspace{0.05\textwidth}
\includegraphics[width=0.45\textwidth]{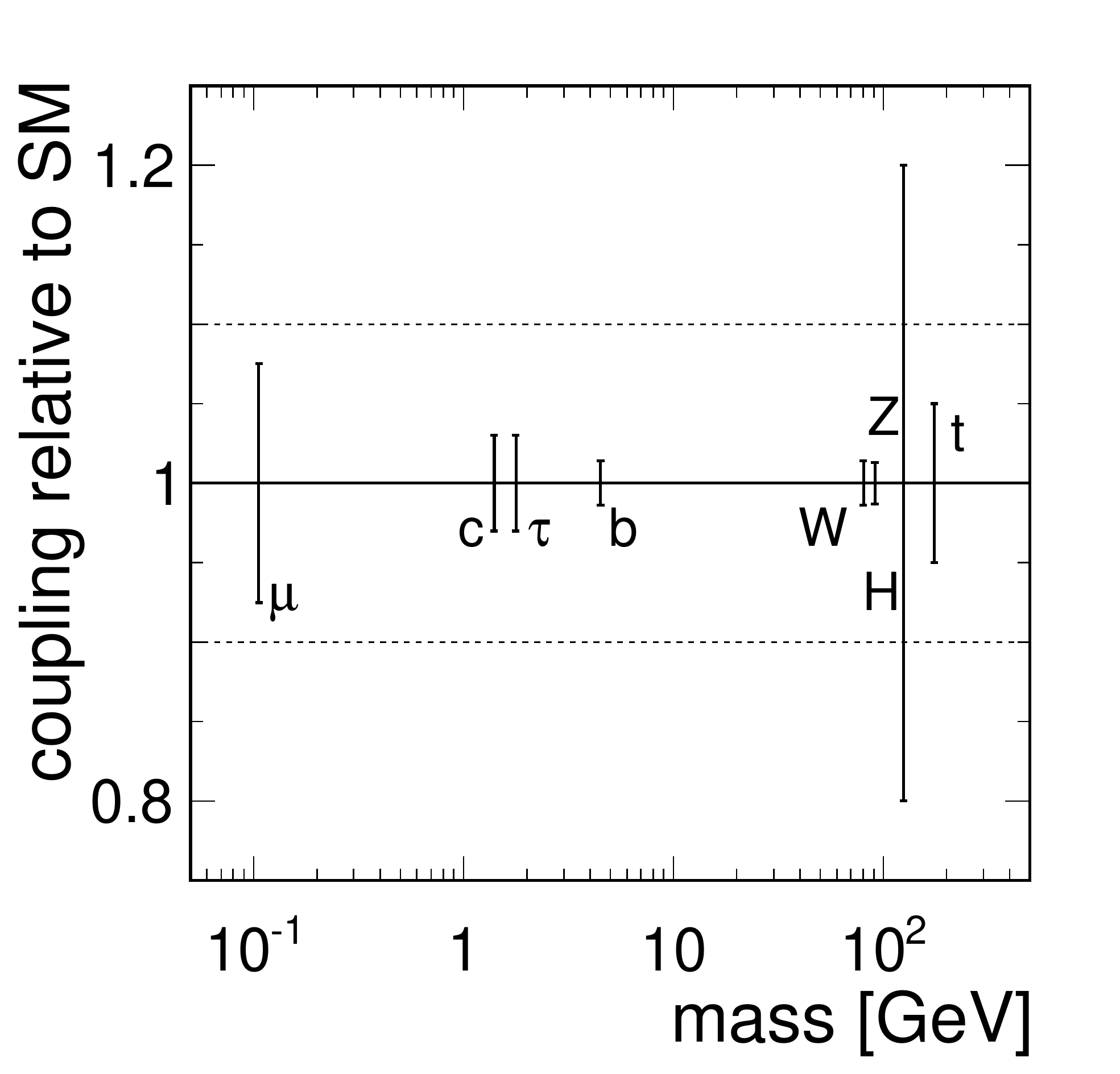}
\caption{Expected precision of couplings to fermions assuming Standard Model couplings proportional to mass (left) and relative precision for coupling measurements to both fermions and bosons (right).}
\label{fig:Couplings}
\end{figure}

Following the discovery of a new boson consistent with the Standard Model Higgs, a linear collider is an excellent option for a comprehensive study of this new form of matter to fully explore the nature of electroweak symmetry breaking. Irrespective of the technology choice for such a collider, its flexibility in energy, further increased by a staged construction, together with the moderate complexity of final states in $e^+e^-$ collisions, provides the prerequisites for a precise measurement of the properties of this particle substantially beyond the capabilities of the LHC. This includes the measurement of the coupling to fermions and bosons, illustrated in Figure \ref{fig:Couplings}, in a model-independent way, the measurement of mass, spin and CP quantum numbers, and direct access to the Higgs potential through the measurement of the trilinear self-coupling. A comprehensive Higgs physics program at a linear $e^+e^-$ collider spans a wide energy range, making full use of the capabilities of the accelerator and of the detectors being developed for this future project.

\end{document}